\documentstyle[preprint,aps]{revtex}
\begin{document}
\preprint{}
\title{
Metastability, Mode Coupling and the Glass Transition}
\author{
Gene F. Mazenko and Joonhyun Yeo}
\address{
The James Franck Institute
and Department of Physics \\
The University of Chicago \\
5640 S. Ellis Avenue,
Chicago, Illinois  60637}
\maketitle
\begin{abstract}
Mode coupling theory (MCT) has been successful in explaining the
observed sequence of time relaxations in dense fluids. Previous expositions
of this theory showing this sequence have required the existence of an
ideal glass transition temperature $T_0$. Recent experiments show
no evidence of $T_0$. We show here how the theory can be reformulated,
in a fundamental way, such that one retains this sequence of relaxation
behaviors but with a smooth temperature dependence and without any
indication of $T_0$. The key ingredient in the reformulation is
the inclusion of the metastable nature of the glass transition problem
through a coupling of the mass density to the defect density. A main
result of our theory is that the exponents governing the sequence of time
relaxations are weak functions of the temperature in contrast to the results
from conventional MCT.
\end{abstract}
\narrowtext
\section{Introduction}
We show here how the condition of metastability in the glass
transition problem can be used to establish the conditions necessary
for a theoretical understanding of the experimentally observed elaborate
sequence of time relaxation behaviors \cite{1} spread over many decades in
time. This time sequence was originally predicted by mode coupling
theory (MCT) \cite{2,3}. We find a smooth temperature dependence for the
relaxation without any evidence of an ideal glass transition temperature as
hypothesized in conventional MCT.

The mode coupling theory for the kinetics near the glass transition
predicts \cite{4} a sequence of time behaviors which has received substantial
verification through a steadily improving series of experiments in
dense liquids \cite{5,6,7}.
This sequence, shown schematically in Fig.1 for the
density-density auto correlation function, begins, after those times
associated with any fast microscopic processes, with a power-law
regime  $f+A_1 t^{-a}$.  For times $t\simeq\tau_a$ there is a
cross over to the so-called von Schweidler regime where the correlation
function decays as $f-A_2 t^b$.  For times $t > \tau_\alpha$ one
moves into the earliest stages of the primary or $\alpha$ relaxation
which can be characterized by a stretched exponential behavior
$A_3 \exp (-(t/\tau )^\beta )$.
There may then be a very long-time crossover
to a final Debye or exponential decay regime.
Within MCT the coefficients
$A_1, A_2$ and $A_3$
are parameters which depend on details of the
model but the exponents $a$ and $b$ are related by a general nonlinear
relationship \cite{4} (discussed below) for which there is some experimental
verification \cite{7}. For certain simplified models \cite{4,8} there is
definite relationship between the exponent $b$ and the stretching exponent
$\beta$. This relationship has been less well studied both theoretically
and experimentally. The prediction of this sequence of time behaviors, which
has substantial experimental support, must be considered a substantial
achievement.

Problems with MCT \cite{9} begin when one considers the temperature dependence
generated by the conventional assumption \cite{4} that there is an ideal glass
transition temperature $T_0$ above and below which the dynamics are
dramatically different. Indeed according to conventional MCT the von
Schweidler and $\alpha$-relaxation regimes are confined to the temperatures
$T>T_0$. This sharp temperature dependence, as pointed out by Kim and
Mazenko \cite{8}, is in disagreement with the high quality experiments of Dixon
et al.~\cite{10}.  Furthermore, and again in disagreement with experiment, the
conventional MCT finds that the sequence of relaxations must be confined to
temperatures near $T_0$ and that the associated exponents $a, b$ and $\beta$
should be temperature independent.  The assumption in conventional MCT is
similar to that associated with critical slowing down near a second order phase
transition.  Experimentally this picture does not hold.  Dixon et al.~\cite{10}
find that the stretching exponent $\beta$ is weakly temperature dependent and
that the qualitative relaxational behavior is the same for all temperatures
spread around any reasonable choice for $T_0$. As pointed out by Kim and
Mazenko, \cite{8} the data of Ref.~10 gives that $( 1 + b ) / ( 1 + \beta )$ is
temperature independent.  Since $\beta$ is temperature dependent the
exponent $b$ must also be temperature dependent.  If the general MCT nonlinear
relationship between $a$ and $b$ \cite{4,21} holds, then $a$ must also be
temperature dependent. MCT theory has even less to say about the
remarkable scaling behavior found by Nagel and coworkers \cite{10}. Kim and
Mazenko \cite{8} have shown that this scaling can be made compatible with MCT
but at the expense of imposing additional ad hoc constraints on the theory.

Our work here is to show how the MCT theory can be reformulated so as to be
compatible with one's intuition that the substantial observed slowing down
results from the fluid system becoming metastable for temperatures below the
melting temperature. The strong temperature dependence in the system goes
into the $\tau$ parameters which characterize the various frequency behaviors.
It appears \cite{11} that these relaxation times show a power-law behavior as a
function of temperature for higher temperatures but one expects such
quantities to show an Arrenhius or Vogel-Fulcher temperature dependence for
sufficiently low temperatures.

Our analysis is based on a model that involves a number of
simplifying assumptions. While most our final results are model dependent,
our final conclusions may be quite general. The model we use is that due to
Das and Mazenko \cite{12},
who developed fluctuating nonlinear hydrodynamics for
the mass density $\rho$ coupled to the momentum density ${\bf g}$, extended to
include a coupling to an additional variable $n$ which we associate with
vacancy diffusion in the system. In the crystal we would be forced to include
this variable, along with the Nambu-Goldstone modes corresponding to
transverse phonons, in a rigorous treatment of the hydrodynamics of such
systems \cite{13}.
Here we assume that it is sensible to define such a variable
in the dense but disordered state \cite{14}. One can think of this in terms of
treating an order parameter in the disordered state.  We will not need a
microscopic definition of this variable. We assume that $n$ is a diffusive
variable with a diffusion coefficient $\Gamma_v$. It is reasonable to
assume that the time scale associated with $n$ is very long, since
one expects that defects must surmount an activation barrier in order to move.
Thus we also assume that the driving effective free energy for $n$ is a
double well potential with the metastable defects associated with the higher
energy well. We assume that the tunneling out of this well is facilitated
by a coupling of $n$ to the mass density in the effective free energy. We
find self-consistently that the stretched dynamics can be associated with a
defect potential with a very small activation barrier and a weak coupling
between $n$ and $\rho$. We show for the model we study that the vanishing of
the barrier and the development of an inflexion point in the potential is a
necessary condition for the slowing down observed experimentally. Apparently
the system must build up the potential in which $n$ moves. We also see that
the notion of a `below' the transition is not well defined from this
point of view.

\section{The Model}
The model we start with is the standard set of equations of fluctuating
nonlinear hydrodynamics for the set of variable $\rho, {\bf g}$ and $n$. We
assume that $\rho$ and $n$ have the usual Poisson brackets of a scalar field
with ${\bf g}$ and that the bare diffusion coefficients are just the bare
shear and bulk viscosities $\eta_0$ and $\zeta_0$
for ${\bf g}$ and a bare diffusion coefficient $\Gamma_v$ for $n$. The
dynamics of this set is then driven by an effective free energy of the form:
\begin{equation}
F=\int\; d^{3} {\bf x}\: \frac{{\bf g}^2 ({\bf x})}{2\rho ({\bf x})} +F_U ,
\end{equation}
where the leading term is the kinetic energy which follows from Galilean
invariance. The term $F_U$ contains the dependence on $\rho$ and $n$
and will be discussed in detail below. The associated Langevin equations of
motion are given then by
\begin{eqnarray}
&& \frac{\partial\rho}{\partial t}= - {\bf \nabla}\cdot {\bf g} \\
&& \frac{\partial g_i}{\partial t}= - \rho\nabla_i \frac{\delta F_U}
{\delta\rho}-n\nabla_i \frac{\delta F_U}{\delta n}-\sum_j\nabla_j
(\rho V_i V_j)-\sum_j L_{ij}V_j + \Theta_i \label{lan:g} \\
&& \frac{\partial n}{\partial t}= - \sum_i \nabla_i (nV_i ) + \Gamma_v
\nabla^2 \frac{\delta F}{\delta n} + \Xi , \label{lan:n}
\end{eqnarray}
where ${\bf V}={\bf g}/ \rho$ and $L_{ij}({\bf x})=-\eta_0 (\frac{1}{3}
\nabla_i\nabla_j +\delta_{ij}\nabla^2 )-\zeta_0\nabla_i\nabla_j$. The bare
longitudinal viscosity is defined by $\Gamma_0 =\zeta_0 +\frac{4}{3}
\eta_0 $. $\Theta_i $ and $\Xi$ in (\ref{lan:g}) and (\ref{lan:n}) are
gaussian noises with variance depending on $L_{ij}$ and $\Gamma_v$.
To complete the specification of our model we must specify the potential
$F_U$. Our simple choice is
\begin{equation}
F_U =\int\; d^3 {\bf x}\: [ \case{A}{2}(\delta\rho)^2+B(\delta\rho)n
+h(n)], \label{FU}
\end{equation}
where $\delta\rho =\rho -\rho_0$ and the defect potential is given by
\begin{equation}
h(n)=\epsilon\bar{n}^4 [\case{1}{4}(\frac{n}{\bar{n}})^4-\case{1}{3}
(2-\sigma )(\frac{n}{\bar{n}})^3+\case{1}{2}(1-\sigma ) (\frac{n}
{\bar{n}})^2\; ],
\end{equation}
where $\bar{n}$ is an average defect density in the metastable well in the
absence of any coupling to $\rho$ and $\epsilon$
gives an over all scale for the
potential and $\sigma$ measures the `distance' from an inflexion point in the
potential, or equivalently, the barrier height. In order to keep the
analysis as simple as possible, we will assume in (\ref{FU})
a very simple form
for the density-dependent part of $F_U$ which corresponds to a
wavenumber independent structure factor. A closely related approximation is
that the correlation function can be factored into a wavenumber dependent part
and a frequency dependent part \cite{19}. This assumption works better than
one might initially guess because, to a first approximation, the slowing down
influences all wavenumbers. This is not a long wavelength or hydrodynamic
approximation.

The way in which one can include the energy variable in the problem
has been discussed by Kim and Mazenko \cite{15}, but we neglect any such
coupling here. The technical problems of treating the nonlinearities
in this model using perturbative field theoretical methods \cite{16} has been
discussed by Das and Mazenko \cite{12}
and clarified by Mazenko and Yeo \cite{17}.
The details of this analysis will be discussed elsewhere \cite{18}. Here
we briefly summarize the nature of the calculation and focus on the results.

For the general set of equations described above with the assumption of the
factorization of wavenumber and frequency dependent parts, one can show that
the Laplace-Fourier transform of the density-density correlation
function can be written in the general form
\begin{equation}
C_{\rho\rho} (z)=\phi (z)=\frac{z+i\Gamma (z)}{z^2-\Omega^2_0
+i\Gamma (z)[z+i\gamma (z)]}, \label{Crr}
\end{equation}
where $\Omega_0$ is a microscopic `phonon' frequency, $\gamma (z)$
is a long-time Das-Mazenko cutoff, and $\Gamma (z)$ is the renormalized or
physical longitudinal viscosity.
Using the methods of Das and Mazenko \cite{12}
one can calculate $\gamma$ and $\Gamma$ in perturbation theory for almost any
choice for the driving free energy. With this background and the
assumption that current correlations decay much faster than $\rho$
correlations,
Das et al.~\cite{20} indicated that
the physical viscosity can be written in the general form
\begin{equation}
\Gamma (z)=\Gamma_0 +\Omega^2_0\int^\infty_0 \; dt \:
e^{izt} H(\phi (t)),
\end{equation}
where the mode coupling kernel can be written in the form
\begin{equation}
H(\phi (t))=\sum^{N}_{n=1} c_n \phi^n (t).
\end{equation}

The MCT analysis is carried out in terms of the coefficients $c_n$. In
the original Leutheussar model \cite{2} one had only the parameter $c_2$.
Das et al.~\cite{20} showed how the development could be generalized to include
an arbitrary set of parameters. We follow here the presentation of Kim and
Mazenko \cite{8}. A significant step forward was made by
G\"otze \cite{21} who realized that if one included a linear
$(c_1 )$ term in $H$ then the von Schweidler and stretching behavior
results. He included such a term on phenomenological grounds. Kim \cite{22}
showed how such a term can be generated by coupling the mass density to an
additional slower variable in the problem. One can also obtain the von
Schweidler form and stretching, and effectively generate a $c_1$ term,
if one includes the wavenumber in the analysis. This has the physical
interpretation that some band of wavenumbers is slower than others and
stretches the faster modes. Some evidence for this picture was found in the
simulations of Valls and Mazenko \cite{23} and the mode coupling calculations
of Fuchs et al.~\cite{24}. Unfortunately an analytical treatment of these
effects are not available. By including the wavenumber dependence in the
development, as in the calculation of Das \cite{25}, one can include such
effects. Here we focus on the wavenumber independent model.

The Das-Mazenko cutoff appearing in (\ref{Crr}) generates an exponential decay
when it kicks in. We assume that the basic mechanism introduced by
Sj\"ogren \cite{26} and discussed by Kim and Mazenko \cite{8} shifts
the values of $\gamma$ to very small values. We assume for the rest of the
discussion here, over the range of time scales we discuss, that this cutoff
can be set to zero.

If there was no coupling of $\rho$ to $n$ then our calculation is
equivalent to that carried out by Das et al.~\cite{20}
and in more detail by Das
and Mazenko \cite{12}
and no linear term in $H(t)$ is produced. The results
in this case are equivalent to the Leutheusser model \cite{2} which has no
stretching and no von Schweidler regime. The basic idea for introducing a
$c_1$ term in the analysis, as discussed by Kim \cite{22} and us
elsewhere \cite{18}, is that the perturbation theory
expansion for $\Gamma$ will generate terms involving quadratic and higher
order products of correlation functions. The
correlation functions involving a current decay faster and those involving
the vacancies slower than the density. Thus the vacancy-vacancy correlation
function $C_{nn}$ can be taken as a constant over the time range where the
density-density correlation function $C_{\rho\rho}$ is stretched \cite{27}.
Thus, for example, the quadratic form $C_{nn}C_{\rho\rho}$ can be
replaced by a constant times $C_{\rho\rho}$ and generates an effective
linear term in $H$.

Metastability enters the problem since there is a significant time
period over which the defects are trapped in a metastable state with
some average density $\bar{n}$. We restrict our perturbation analysis to
this time regime. A key aspect of our analysis is that because $n$ is
coupled to $\rho$ the stationary value of $\bar{n}$ depends on the local value
of $\rho$ determined by looking for the metastable minimum of $F$ which is
given by
\begin{equation}
n^{*}=\bar{n} [1-\frac{x}{\sigma y} (\frac{\delta\rho }{\rho_0})
-(\frac{1+\sigma}{\sigma})(\frac{x}{\sigma y})^2
(\frac{\delta\rho }{\rho_0})^2+\cdots ]
\end{equation}
where we have the dimensionless parameters
$x=B\rho_0\bar{n}/A\rho^2_0$
(a measure of the coupling between $\rho$ and $n$) and
$y=\epsilon\bar{n}^4 /A\rho^2_0 $.
Our procedure
is then to expand $n$ about $n^{*}$ in the Langevin equations and then carry
out perturbation theory in the nonlinear terms keeping terms up to $N=2$ in
the mode coupling kernel. Self-consistently we find substantial slowing of
the dynamics in the cases where $x$ and $\sigma$ are very small. The smallness
of $\sigma$ corresponds to a low barrier in $h(n)$ while small $x$ means
that the potential is not greatly distorted by fluctuations in $\rho$. The
limit in which we obtain self-consistency is where $x=C\sigma^2$ with
$C$ of 0(1). We then obtain, eliminating $x$ in terms of $C$ and $\sigma$
for small $\sigma$, that
\begin{eqnarray}
&&c_1=\xi [-8(y+C)-4(\case{C^2}{y}+3y+8C)\sigma ]+ \mbox{O}
(\sigma^2 ) \\
&&c_2=\xi [1+4y+2(3y+2C)\sigma ]+\mbox{O}(\sigma^2 ),
\end{eqnarray}
where the formal expansion parameter
$\xi=k_B T\Lambda^3/(6\pi^2 A\rho^2_0 )$
and $\Lambda$ is a short-distance cut-off.
The terms $c_3, c_4, \cdots$ are available if one goes to higher order
in perturbation theory.

We now demand that the parameters characterizing the defect well and the
coupling are chosen to give the slow dynamics observed experimentally. This
involves the mathematical machinery developed by MCT \cite{4,21}
but interpreted
differently \cite{8}.
This analysis can be stated rather generally. If we define
\begin{eqnarray}
&& \sigma_0 =(1-f)V(f) \\
&& \sigma_1 =(1-f)^2 V^{\prime}(f) \\
&&\lambda =\case{1}{2}(1-f)^3 H^{\prime\prime}(f),
\end{eqnarray}
where $V(f)=H(f)-f/(1-f)$ and $f$ has the physical interpretation
as the metastable value of $\phi$ evaluated for the critical conditions
$\sigma_0=\sigma_1=0$ which characterize the slow dynamics. In
this case it can be shown that the decay exponents
$ a$ and $b$ are given by \cite{21}
\begin{equation}
\frac{\Gamma^2 (1-a)}{\Gamma (1-2a)}=\lambda =
\frac{\Gamma^2 (1+b)}{\Gamma (1+2b)}
\end{equation}
In the present case where we include only $c_1$ and $c_2$ the
critical conditions $\sigma_1=0$ can be used to eliminate the parameter
$f$ in equations (13-15) to obtain for small $\sigma_0$,
\begin{eqnarray}
&&c_1 =\frac{2\lambda -1}{\lambda^2} + \frac{4\sigma_0}
{\lambda (1-\lambda )}+ \cdots \\
&&c_2 =\frac{1}{\lambda^2}-\frac{3\sigma_0}
{\lambda(1-\lambda )^2}+\cdots
\end{eqnarray}
Comparing (17) and (18) with (11) and (12) we obtain, assuming
$\sigma_0\simeq\sigma$,
four equations which can be solved to give $C,y$ and $\lambda$ as
functions of $\xi$ given by
\begin{eqnarray}
&&C=\case{1}{4}(1-\case{2\lambda+1}{2\xi\lambda^2}) \\
&&y=\case{1}{4}(\case{1}{\xi\lambda^2}-1)
\end{eqnarray}
and we have the implicit equation for $\lambda$ as a function of
$\xi$ given by
\begin{equation}
\xi=\frac{1}{\lambda^2}\{ 1-\frac{3(2\lambda -1)}{
2[7+2\lambda +\sqrt{4\lambda^2 +22\lambda +91}]   }   \}.
\end{equation}
We can also express $\sigma$ in terms of $\sigma_0$ and $\xi$:
\begin{equation}
\sigma=-\frac{3}{2[3y+2C]\xi\lambda(1-\lambda )^2}
\sigma_0 .
\end{equation}
For $\frac{1}{2}<\lambda <1$, $\sigma$ and $\sigma_0$ have the same sign.
However, as can be seen from (6), only the absolute value $| \sigma |$ matters
in determining the critical condition, since the potential $h(n)$ is
`symmetric' \cite{28} around $\sigma=0$ in the sense that both $\sigma >0$
and $\sigma <0$ cases represent a double-well potential.

\section{Discussion and Conclusion}
Using the results of the previous section, we can express the exponents $a$
and $b$ and the amplitude $f$
as weak functions of temperature, $\xi$, as shown in Fig. 2. Note that we
have been able, except for the temperature dependence of $\sigma$, to determine
the form of the potential $h ( n )$ and its coupling to $\rho$ as a function of
temperature as represented by the parameter $\xi$.  At this stage it looks
worthwhile to try and determine $\sigma$ experimentally.

Our analysis here has focused on the time regime prior to the
$\alpha$-relaxation and we have not yet addressed the question of the scaling
found by Dixon et al.~\cite{10}. Our detailed results giving $a , b$
and $f$ as
functions of $\xi$ are model dependent and since our model is over-simplified
should not be expected to apply to experiments in detail. However it is
possible that our basic picture that $a , b$ and $f$ are weak functions of
temperature and that the parameter $\sigma$ controls the slowing down may prove
to be robust.

\section*{
Acknowledgements}

This work was supported by the National Science Foundation Materials
Research Laboratory at the University of Chicago.

\begin{figure}
\caption{
A schematic plot of the sequence of relaxation behaviors predicted
by MCT; (a) power-law decay relaxation $f+A_1 t^{-a}$; (b)
von-Schweidler relaxation $f-A_2 t^b$; (c) primary relaxation
$A_3 e^{-(t/\tau )^\beta }$; and (d) exponential relaxation
$e^{-\gamma t}$.
}
\end{figure}
\begin{figure}
\caption{
The exponent parameters $a$, $b$ and the metastability parameter
$f$ as functions of the temperature represented by $\xi$. The
condition $\frac{1}{2}<\lambda <1$ restricts the range of $\xi$
to $0.93< \xi <4$.
}
\end{figure}


\begin{references}
\bibitem{27}
This separation of times scales between the density and the vacancy
variables result if the coupling $b$ between $\rho$ and $n$ and $\sigma$ are
very small and satisfy a certain limit which will be obtained in the following
discussion. In this limit, the density-vacancy correlation function
$C_{\rho n}$ is a linear combination of $C_{\rho \rho}$ and $C_{nn}$.
\bibitem{28}
The form of the potential $h ( n )$ is invariant under
$\bar{n}\rightarrow\bar{n}^{\prime}
\equiv (1-\sigma )\bar{n}$
and $\sigma\rightarrow\sigma^{\prime}\equiv -\sigma/
(1-\sigma )$.
Thus $h ( n )$ with $\bar{n}$ and $\sigma <0$ is the same double-well potential
as the one with $\bar{n}^{\prime}$ and $\sigma^{\prime} > 0$.
\end{references}
\end{document}